\documentclass[a4paper,11pt]{article}
\usepackage[export]{adjustbox}
\usepackage[english]{babel}
\usepackage[utf8]{inputenc}
\usepackage{pos}
\usepackage{amsmath}
\usepackage{color,graphicx}
\usepackage{bm,bbm,braket}
\usepackage{comment}
\usepackage{enumerate}
\usepackage{hyperref}
\usepackage{multirow}
\usepackage{mathtools}
\usepackage{mathrsfs}
\usepackage{mciteplus}
\usepackage{slashed}
\usepackage{soul}
\usepackage{subfigure}
\usepackage{wrapfig}
\usepackage{xspace}

\renewcommand{\k}{\bm{k}}

\newcommand{\3}{\bm{3}}

\newcommand{\Cc}{\mathcal{C}}

\newcommand{\Kc}{\mathcal{K}}

\newcommand{\Mc}{\mathcal{M}}

\newcommand{\Rc}{\mathcal{R}}

\DeclareMathOperator{\re}{Re}
\DeclareMathOperator{\im}{Im}

\newcommand{\bcol}{\left[ \begin{array}{c}}
\newcommand{\ecol}{\end{array} \right]}
\newcommand{\beq}{\begin{eqnarray}}
\newcommand{\eeq}{\end{eqnarray}}

\newcommand{\kev}{\ensuremath{{\mathrm{\,ke\kern -0.1em V}}}\xspace}
\newcommand{\mev}{\ensuremath{{\mathrm{\,Me\kern -0.1em V}}}\xspace}
\newcommand{\gev}{\ensuremath{{\mathrm{\,Ge\kern -0.1em V}}}\xspace}
\newcommand{\tev}{\ensuremath{{\mathrm{\,Te\kern -0.1em V}}}\xspace}

\usepackage{mfirstuc} 
\newcommand{\addReviewer}[2]{
  \expandafter\newcommand\csname #1\endcsname[1]{{\bf \color{#2} \capitalisewords{#1}:\,##1}}
  \expandafter\newcommand\csname #1cor\endcsname[2]{{\color{#2} \capitalisewords{#1}:\,\st{##1}{\bf ##2}}}
  \expandafter\newcommand\csname #1color\endcsname{#2}
}

\usepackage{soul,color}
\definecolor{chromeyellow}{rgb}{1.0, 0.65, 0.0}
\definecolor{DodgeBlue}{rgb}{0.118, 0.565,1.000}
\definecolor{asparagus}{rgb}{0.53, 0.66, 0.42}
\definecolor{cardinal}{rgb}{0.77, 0.12, 0.23}
\definecolor{cadmiumgreen}{rgb}{0.0, 0.42, 0.24}
\definecolor{applegreen}{rgb}{0.55, 0.71, 0.0}

\addReviewer{wrong}{red}
\addReviewer{good}{green}
\addReviewer{sebastian}{cardinal}

\title{Analytic continuation of the finite-volume three-particle amplitudes}

\author*[a]{Sebastian M. Dawid}

\affiliation[a]{Physics Department, University of Washington, Seattle, WA 98195-1560, USA}

\emailAdd{dawids@uw.edu}

\abstract{One has to study multivariable scattering amplitudes to extract properties of the three-body states from the generalizations of the L\"uscher finite-volume formalism. In particular, a three-body amplitude obtained from a Lattice QCD calculation must be analytically continued to unphysical Riemann sheets of the complex energy plane, where resonances of interest appear as poles. In this article, we provide a pedagogical overview of a method for solving and analytically continuing the on-shell integral equations describing a three-body elastic scattering process. We illustrate the procedure by applying it to a relativistic system of three identical bosons characterized by pair-wise interactions. We describe the analytic structure of the reaction amplitude, show how to access its physical and unphysical Riemann sheets, and analyze the behavior of the three-body spectrum under variations of the interaction parameters.}

\FullConference{The 40th International Symposium on Lattice Field Theory (Lattice 2023)\\
July 31st - August 4th, 2023\\
Fermi National Accelerator Laboratory\\}

\begin{document}
\maketitle

\section{Introduction}
\label{sec:Introduction}

The ongoing three-body spectroscopy program~\cite{Horz:2019rrn, Blanton:2019vdk, Mai:2019fba, Culver:2019vvu, Fischer:2020jzp, Hansen:2020otl, Alexandru:2020xqf, Brett:2021wyd, Blanton:2021llb, Draper:2023boj} aims to use the lattice formulation of QCD to precisely determine properties of hadronic resonances characterized by significant three-particle decay channels. These include curious states like $\omega(782)$, $\pi_1(1600)$, or $T_{cc}^+(3872)$, and many more. One study such hadrons with the generalization of the L\"uscher's formula~\citep{Luscher:1985dn, Luscher:1986pf}, known as the three-body quantization condition~\cite{Hansen:2014eka, Hansen:2015zga, Briceno:2018aml, Blanton:2020gha}. Analogously to the L\"uscher's formalism, it allows one to translate the spectrum of the finite-volume theory into the $\3 \to \3$ scattering amplitude, $\Mc_3$. Since resonant states manifest as complex poles of $\Mc_3$, one has to analytically continue this function to complex energies to extract their masses and widths.

Such a procedure has been well established as a required supplement to the quantization condition in the two-body analyses of the lattice data; see Refs.~\citep{Rodas:2023gma, Wilson:2023anv} for the recent advanced examples. Rigorous understanding of the amplitude's analytic properties became obligatory when determining a model-independent interpretation of states appearing in realistic scattering systems~\citep{Green:2021qol, Korpa:2022voo, Du:2023hlu, Meng:2023bmz, Raposo:2023oru}. This necessity is even more pressing when describing the $\3 \to \3$ reactions, which require a solution to a system of integral equations~\cite{Hansen:2015zga, Blankenbecler:1965gx, Aaron:1968aoz, Mai:2017vot, Jackura:2018xnx, Jackura:2020bsk} containing singular, multi-variable functions of the particles momenta~\cite{Hwa:1964xyz, Rubin:1966zz, Rubin:1967zza}. 

In this article, we desire to present a general procedure for the analytic continuation of these relativistic three-body integral equations, as developed and discussed in Refs.~\citep{Dawid:2023jrj, Dawid:2023kxu}. These two works extend and clarify techniques introduced and studied by Hetherington and Schick~\citep{PhysRev.137.B935}, Brayshaw~\cite{PhysRev.167.1505, Brayshaw:1968yia} and Gl\"ockle~\cite{Glockle:1978zz} in the non-relativistic context. The procedure is applicable in the infinite-volume analyses of the three-body lattice spectrum. We start with a short review of the infinite-volume counterpart of the three-body scattering formalism of Ref.~\citep{Hansen:2015zga} in Sec.~\ref{sec:Integral-equations}. In Sec.~\ref{sec:analytic_continuation}, we discuss the analytic continuation of the amplitude defined via the Lippmann–Schwinger-like integral equation. We try to keep the discussion simple and conceptual, hoping to make it a useful introduction to Ref.~\citep{Dawid:2023jrj}. In Sec.~\ref{sec:results}, we present the results of applying this procedure to the simple $\3 \to \3$ process. We shortly discuss curious properties of the spectrum unique to the three-body physics—such as the cyclic trajectories of resonances and the Efimov phenomenon. We close the article in Sec.~\ref{sec:Conclusions} with a summary.

\section{Relativistic three-body scattering equation}
\label{sec:Integral-equations}

Following Refs.~\citep{Jackura:2020bsk, Dawid:2023jrj, Dawid:2023kxu} we study the connected three-body amplitude $\Mc_{3}$ within a framework of a generic relativistic scalar EFT~\cite{Hansen:2015zga}. The amplitude describes the probability for elastic scattering of three identical spin-zero particles of mass $m$, denoted here by $\varphi$. For simplicity, we consider exclusively the $S$-wave reaction, i.e., a single partial-wave component of $\Mc_3$. As discussed in Ref.~\citep{Dawid:2023jrj}, it does not affect the generality of our procedure, which applies to sectors of higher angular momentum. To further simplify the description of the problem, we make a stronger assumption that the short-range three-body couplings, described jointly by the so-called three-body $K$ matrix, $\Kc_{\text{df},3}$, vanish~\citep{Hansen:2014eka, Romero-Lopez:2019qrt, Jackura:2020bsk}. The scattering process becomes generated solely by repeated on-shell particle exchanges (OPEs). We note that it is likely necessary to include a non-zero $K$ matrix to accurately describe three-body states occurring in nature. However, a large enough class of possible $K$'s does not substantially affect the method of analytic continuation described below. It justifies proceeding with the simplified model.

As for the kinematics of the reaction, the particles interact with the total invariant mass squared $s=E^2$, where $E$ is the energy in their c.m.~frame. Furthermore, we divide a three-particle state into a single particle (a \emph{spectator}) and remaining particles in a corresponding two-body subchannel (a \emph{pair}).\footnote{In the language of Ref.~\cite{Hansen:2015zga}, we always investigate the ``unsymmetrized" amplitude $\Mc^{(u,u)}_3$ of a quasi two-body pair-spectator process.} In the total c.m. frame, a spectator of momentum $k = |\k|$ has energy $\omega_k = \sqrt{k^2 + m^2}$. The corresponding pair is characterized by the total invariant mass squared, $\sigma_k = (\sqrt{s} - \omega_k)^2 - k^2$. We use $s$ and initial and final spectator momenta, $k$ and $p$, as the independent variables describing the scattering. 

In this setting, one obtains the amputated partial-wave projected amplitude $d(p,k)$,
    \begin{align}
    \Mc_3(p,k) \equiv \Mc_2(p) \, d(p,k) \, \Mc_2(k)
    \end{align}
by solving the (\emph{ladder}) integral equation,
    \begin{align}
    \label{eq:Ddef_v2}
    d(p,k)
    = 
    - G(p,k) 
    - \int\limits_0^{k_{\rm max}} \! \frac{{\rm d} k' \, k'^2}{(2 \pi)^2 \omega_{k'}} 
    \, G(p,k') 
    \, \Mc_2(k') 
    \, d(k',k)  \, ,
    \end{align}
where the implicit $s$ dependence is assumed~\citep{Hansen:2015zga}. We use the letter $d$ for the amplitude to distinguish it from the case of $\Kc_{\text{df},3} \neq 0$.\footnote{A non-zero three-body $K$ matrix leads to an additional contribution to $\Mc_3$, not discussed in this article.} The equation is similar to the Lippmann–Schwinger equation for the off-shell two-body $T$ matrix (assuming $G$ is the potential and $\Mc_2$ is the energy denominator) and can be solved numerically using comparable methods~\citep{Jackura:2020bsk}. However, Eq.~\eqref{eq:Ddef_v2} describes an on-shell amplitude for a pair-spectator reaction, where $G$ governs on-shell particle exchanges while $\Mc_2$ on-shell subchannel processes.  

Before defining these objects, let us note that $d(p,k)$ can have complex poles in the variable $s$ corresponding to three-body states. The amplitude has branch cuts associated with the open scattering channels, and these poles can be found either on the first Riemann sheet (three-body bound states) or the nearest unphysical sheet (virtual states and resonances) corresponding to these discontinuities. The problem of identifying states of interest is non-trivial given that the objects building the ladder equation, when viewed as complex functions, have various singularities in all three variables, $p,k$, and $s$. We wish to explain their nature, describe the resulting discontinuities of $d$, and extend the applicability of Eq.~\eqref{eq:Ddef_v2} to arbitrary complex energies on various Riemann sheets.

In the ladder equation, $\Mc_2$ is the $S$-wave $2\varphi \to 2 \varphi$ amplitude, implied by the nature of interactions between particles in the pair. For concreteness, in the following, we use a low-energy approximation,
    \beq
    \label{eq:M2_general}
    \Mc_2(k') = \frac{16 \pi \sqrt{\sigma_{k'}}}{-1/a - i \sqrt{\sigma_{k'} / 4 - m^2} } \, ,
    \eeq
where $a$ is the two-body scattering length and the only parameter of the model. This two-body amplitude has a pole at $k' = q = \lambda^{1/2}(s,m^2,m_b^2)/2\sqrt{s}$, where $m_b^2 = 4\left(m^2-1/a^2 \right)$ and $\lambda(x,y,z) = x^2 + y^2 + z^2 - 2(xy+yz+zx)$. Moreover, $\Mc_2(k')$ has branch points at the threshold, $k' = k_{r} = \lambda^{1/2}(s, m^2, (2m)^2)/2 \sqrt{s}$, and pseudo-threshold, $k' = k_{l} = \lambda^{1/2}(s, m^2, 0)/2 \sqrt{s}$. We orient the associated cuts to run to the complex infinity. We note that all these singularities in $k'$ are ``movable", i.e., are functions of another parameter, $s$. Moreover, they have mirror copies in the complex $k'$ plane obtained by their reflection with respect to the point $k'=0$. 

The presence of the branch points implies that the two-body amplitude has multiple Riemann sheets associated with the listed cuts. For instance, the second-sheet value of $\Mc_2(k')$, corresponding to the cut starting at $k_r$, is,
    \beq
    \label{eq:M2_2nd}
    \Mc_2^{\text{II}}(k') = \frac{16 \pi \sqrt{\sigma_{k'}}}{-1/a + i \sqrt{\sigma_{k'} / 4 - m^2} } \, .
    \eeq

The second building block of Eq.~\eqref{eq:Ddef_v2} is the $S$-wave-projected OPE amplitude, $G$, which governs a probability for a boson exchange between the incoming and outgoing pairs,
    \begin{align}
    \label{eq:Gs_proj}
    G(p, k)
    & = - \frac{H(p, k)}{4pk} \, \log\left( \frac{z(p, k) - 2pk + i \epsilon }{z(p, k) + 2pk + i \epsilon } \right) \, .
    \end{align}
The function $z(p, k) = (\sqrt{s}-\omega_{k} - \omega_p)^2 - k^2 - p^2 - m^2$. Quadratic momentum dependence in the argument of the logarithm leads to a convoluted formula~\citep{Brayshaw:1968yia, Glockle:1978zz, Orlov:1984, Sadasivan:2021emk} parameterizing the cut of $G(p,k)$ in the complex $p$ plane,
    \beq
    \label{eq:OPE-discontinuity}
    p_{+}(s,k,x) = \frac{1}{2\beta_x} \Big( k x \, (\beta_1 + i \epsilon) + \sqrt{\beta_0} \sqrt{(\beta_1 + i \epsilon)^2 - 4 m^2 \beta_x} \Big) \, .
    \eeq
Here we introduced an auxiliary function, $\beta_x = (\sqrt{s} - \omega_k)^2 - x^2 k^2 $ and a real parameter $x \in [-1,1]$. The cut runs between two branch points reproduced by Eq.~\eqref{eq:OPE-discontinuity} for $x=\pm 1$. Similarly to singularities of $\Mc_2$, it has a mirror copy given by $-p_+$. 

The factor $H(p,k)$ in Eq.~\eqref{eq:Gs_proj} describes a regularization choice~\citep{Hansen:2014eka}. We consider two types of cut-offs represented by a smooth or discontinuous (``hard" cut-off) function $H(p,k)$~\citep{Dawid:2023jrj}. In both cases, the upper limit of the integration is $k_{\rm max} = k_l$, i.e., it does not exceed the value of the pseudo-threshold of $\Mc_2$. It ensures that the pseudo-threshold cut of the two-body amplitude does not affect the analytic properties of $d(p,k)$.

It is the presence of the logarithmic singularity in Eq.~\eqref{eq:Gs_proj} that makes the solution and analytic continuation of Eq.~\eqref{eq:Ddef_v2} non-trivial, as known from the beginnings of the quantum-mechanical three-body problem~\citep{Skorniakov:1957kgi, belyaev1990lectures}. In particular, for some values of external momenta and total invariant mass, the branch points of the OPE coincide with the integration interval, requiring its generalization to a complex deformed contour. In the most complicated case, both mirror copies of the branch cut can merge and form the so-called circular cut surrounding the integration endpoint. Although the position of the branch cuts is a matter of convention which, in principle, could be altered to produce a simpler geometry of these singularities, in practice, such redefinitions typically lead to equivalent or additional complications. Below, we describe how to deal with them.

\section{Analytic continuation of the ladder equation}
\label{sec:analytic_continuation}

\subsection{General idea}

It is insufficient to generalize $s$ from a real to a complex variable and then solve the integral equation (numerically) to obtain the amplitude for an arbitrary energy. As mentioned in the previous section, it happens that for specific values of $s$, singularities of the integration kernel $K(p,k') = -G(p,k') \Mc_2(k')$ cross the integration interval, invalidating simplistic extrapolation of the three-body equation to complex energies.

Forgetting for a moment about the inhomogeneous term of the ladder equation, we can view the amplitude $d(s) = d(p,k)$ as a complex function defined as an integral of another (unknown) function,
    \beq
    \label{eq:simple_d}
    d(s) =  \!\!\!\int\limits_{\Cc(0,k_{\text{max}})} \!\!\!\!\!\! f(k', s) \, dk' \, + \, \dots \, .
    \eeq
Here $f(k', s) = K(p,k') d(k', k)$, where we highlight the dependence of the integrand on the argument $k'$ and a complex parameter $s$ and keep the $(p,k)$ dependence implicit. The integration contour, $\Cc(0,k_{\text{max}})$ starts at $k' = 0$ and ends at $k' = k_{\text{max}}$. 

Although we do not know $f(k',s)$ beforehand, we know the positions of its singularities in $k'$ and how they change with $s$. It is enough to infer the analytic structure of $d(s)$ in the complex $s$ plane~\cite{Eden:1966dnq, burkhardt1969dispersion}. In particular, its singularities are present not only when the integrand exhibits an explicit non-analyticity in $s$ but also when an $s$-dependent (movable) singularity in the $k'$ variable coincides with the lower endpoint of the integration contour, $k'=0$.

Let us illustrate it with a concrete example. The $d$ amplitude has two branch points corresponding to two reaction thresholds, implied by the $S$ matrix unitarity. One occurs at $s_{\varphi b} = (m+m_b)^2$ and is associated with a physical $\varphi$ particle scattering off the $2\varphi$ bound-state, denoted here by $b$. It is present only for $a>0$, i.e., when amplitude in Eq.~\eqref{eq:M2_general} develops a bound-state pole. The branch cut is algebraic in nature and has two associated Riemann sheets. The other branch point, at $s_{3\varphi} = (3m)^2$, is related to the possibility of a physical $3 \varphi \to 3 \varphi$ process and is independent of $a$. It is logarithmic and has an infinite but countable number of Riemann sheets~\citep{LANDAU1959181, Eden:1966dnq}. The origin of these two singularities of $d(s)$, in the context of Eq.~\eqref{eq:simple_d}, is as follows. The integrand, $f(k',s)$, has a pole at $k'= q$, which travels in the complex $k'$ plane with varying $s$. The pole collides with the point $k'=0$ at $s = s_{\varphi b}$, as seen from Eq.~\eqref{eq:M2_general}. On the other hand, the three-body cut is born from the collision of the two-body amplitude's branch point at $k' = k_r$ with $k'=0$ at energy $s = s_{3 \varphi}$. 

These two structures are independent of $G(p,k')$. However, in addition to the two threshold singularities required by unitarity, the amplitude develops an unphysical singularity from the collision of the OPE branch point $p_+$ with $k'=0$. We do not discuss it here and refer the reader to Ref.~\citep{Dawid:2023jrj} for more details. Finally, for each $(p,k)$, the three-body amplitude inherits an explicit logarithmic cut from the OPE amplitude in the first term of Eq.~\eqref{eq:Ddef_v2}, so far neglected in this analysis.

\subsection{Appropriate integration contour}

Whenever singularities of the kernel, $q, k_r, p_+$, cross the integration interval (but not the endpoint), one can avoid integrating over them by deforming the integration path. As inferred by the Cauchy theorem, this allows one to analytically continue $d(s)$ to values of $s$ where the na\"ive extrapolation is ill-defined. The shape of the contour (i.e., the direction from which it circumvents the poles and branch points) defines the Riemann sheet at which one probes $d(s)$.

The deformed integration contour must belong to the region of the complex plane where the function $f(k',s)$ is analytic. It imposes a set of restrictions on the acceptable paths $\Cc(0,k_{\text{max}})$. The most strict constraint is imposed by the need to avoid the logarithmic cut of Eq.~\eqref{eq:OPE-discontinuity}. This cut is present not only in $G(p,k')$ but also in $d(k',s)$, which acquires it from both terms on the right-hand side of Eq.~\eqref{eq:Ddef_v2}.

Focusing on the first one, for a fixed choice of $s$ and $k$, $G(p,k)$ contributes the logarithmic cut to the $p$ dependence of $d(p,k)$. One avoids it by an appropriate choice of the contour $\Cc$ in the homogeneous term of the integral equation. In the second term of Eq.~\eqref{eq:Ddef_v2}, $G(p,k')$ contributes to $d(p,k)$ one cut in the complex $p$ plane for each value of $k' \in \Cc$. Altogether, they cover a region in the complex $p$ plane where $d(p,k)$ is non-analytic. This region, referred to as the \emph{domain of non-analyticity}, $\bar{\Rc}_{\Cc}$, depends on the contour introduced in the earlier step. The integration path must detour $\bar{\Rc}_{\Cc}$ that it defines, i.e., it has to be \textit{self-consistent}. We refer the reader to Fig.~8 in Ref.~\citep{Dawid:2023jrj} for an illustration of a typical domain of non-analyticity. Once we ensure that the chosen integration path satisfies the self-consistency criterion, the integral equation can be solved numerically along the complex integration path.

\begin{figure}[b]
    \centering
    \includegraphics[width=0.8\textwidth]{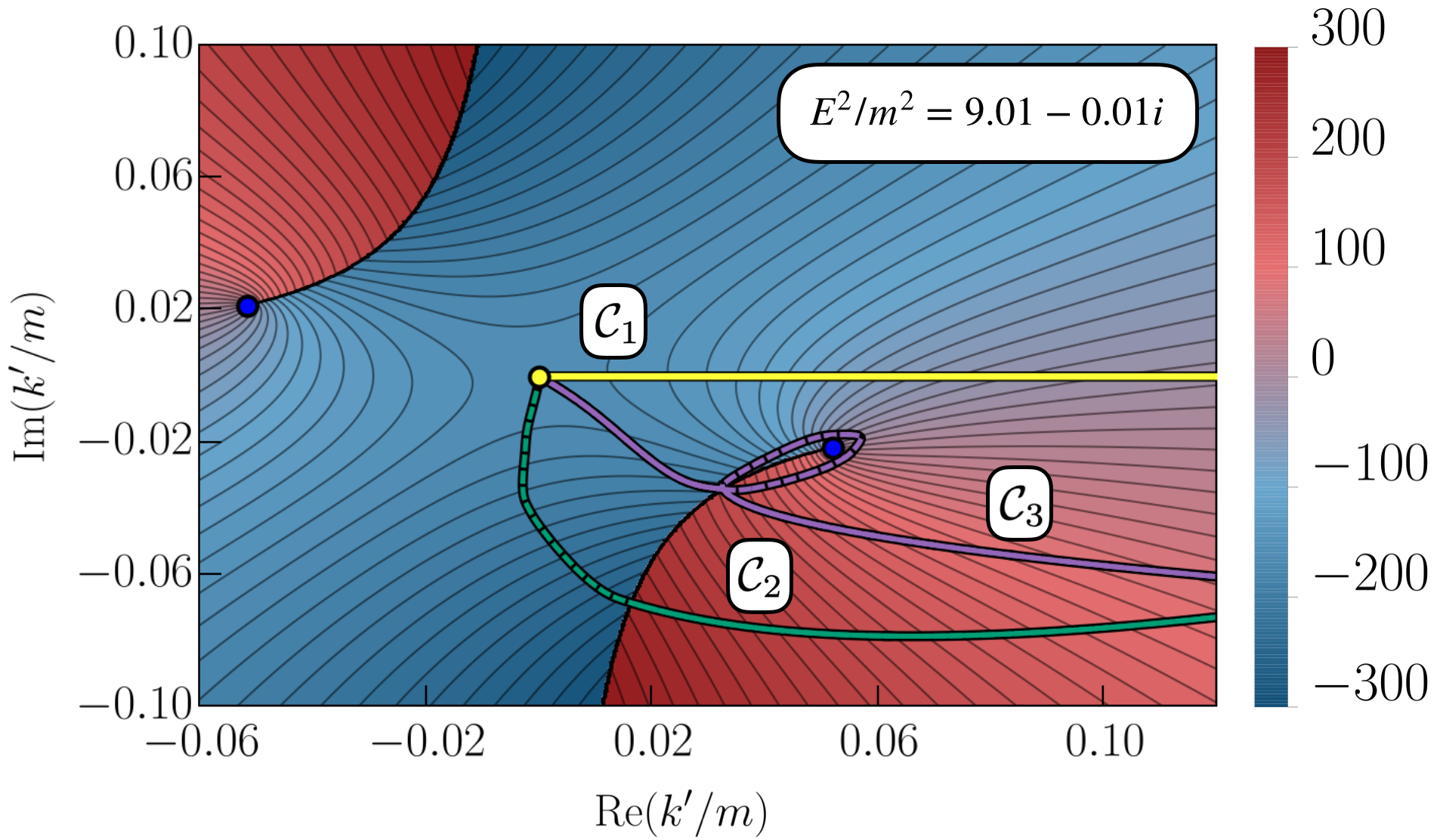}
    \caption{$\im \Mc_2(k')$ present in the integration kernel $K(p,k')$. Blue dots highlight branch points at $k' = \pm k_r$. Contour $\Cc_1$ (yellow) defines the solution, $d(p,k)$, on the physical Riemann sheet in the $E^2$ variable. Contour $\Cc_2$ (green) allows one to access the nearest (unphysical) sheet. Contour $\Cc_3$ (purple) defines the amplitude on the next Riemann sheet. Parameters $ma=-6$ and $(E/m)^2 = 9.01 - 0.01i$. Adaptation of Fig.~5 from Ref.~\citep{Dawid:2023kxu}.}
    \label{fig:1}
\end{figure}

\subsection{Continuation through the $\varphi b$ and $3\varphi$ cuts}

As mentioned, one chooses between different self-consistent contours to access unphysical Riemann sheets of $d(p,k)$ associated with the bound-state–particle and three-body thresholds. When $s$ becomes complex, the $\Mc_2$ pole at $k' = q$ and the branch point at $k' = k_{r}$ travel off the real $k'$ axis. The integration contour $\Cc$ running from $k' = 0$ to $k' = k_{\text{max}}$ can bypass the pole or branch point from the top or bottom. By Cauchy's theorem, these two choices lead to solutions different by the value of the loop integral of $f(k',s)$ around the singularity. These different values define distinct Riemann sheets of the three-body amplitude. Formally, this defines the discontinuity and Riemann sheets of the solution to the integral equation via its monodromy~\citep{Bourjaily:2020wvq}.

We present an example of this situation in Fig.~\ref{fig:1}. Three integration contours correspond to the amplitude $d(p,k)$ on three different Riemann sheets associated with the three-body threshold. Deformation from the straight line, $\Cc_1$, to a complex contour $\Cc_2$, takes one from the physical sheet to the first unphysical one. To maintain the continuity of the integration kernel, the 2nd sheet value of $\Mc_2(k')$, given in Eq.~\eqref{eq:M2_2nd}, is used when $k'$ belongs to the dashed portion of $\Cc_2$ (i.e., before crossing the cut). Finally, one employs integration contours that encircle the branch point to obtain  $d(p,k)$ on higher sheets. For instance, contour $\Cc_3$ allows one to evaluate the solution on the subsequent Riemann sheet. Again, when the integration variable belongs to the dashed part of the contour, $\Mc_2^{\text{II}}$ is used to ensure analyticity. For completeness, Sec.~IV.E of Ref.~\citep{Dawid:2023jrj} discusses an example of continuation beyond the $\varphi b$ branch cut.

\section{The three-body spectrum}
\label{sec:results}

\begin{figure}[b]
    \centering
    \includegraphics[width=0.99\textwidth]{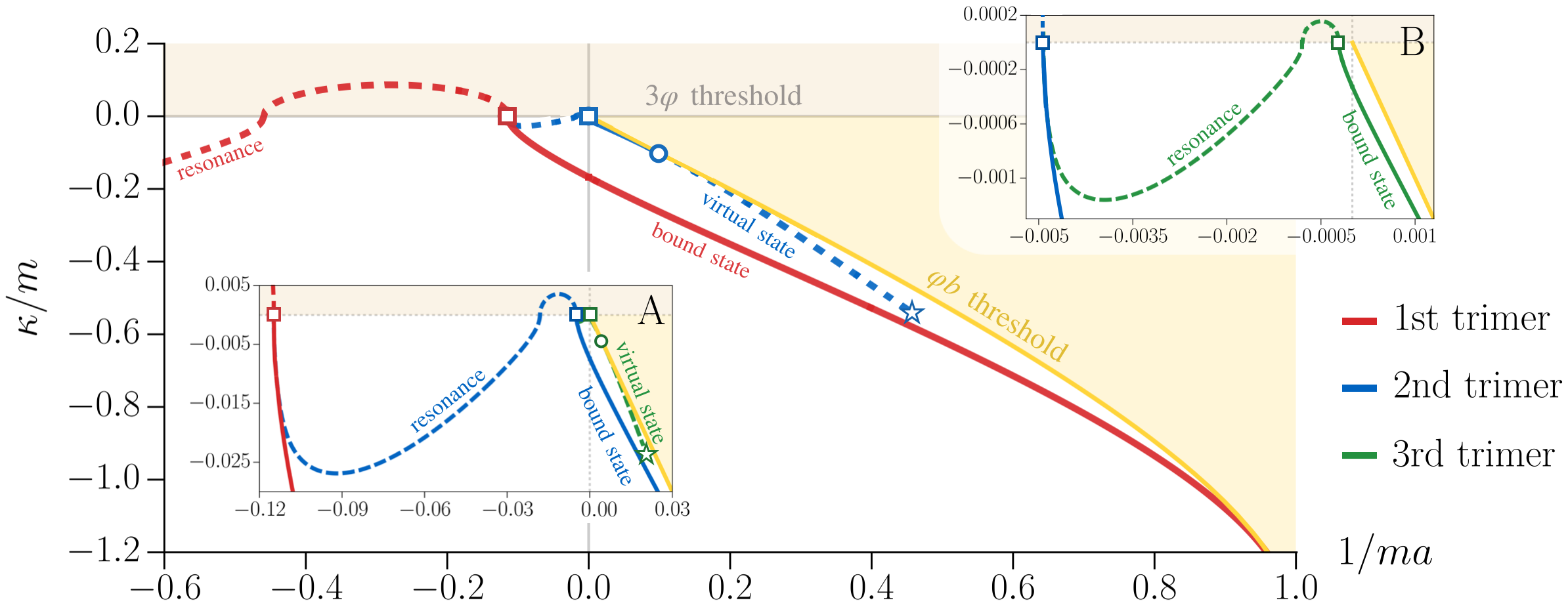}
    \caption{First three levels of the three-body spectrum. Solid lines denote bound states, dashed lines either virtual states on the unphysical $\varphi b$ sheet or resonances on the unphysical $3\varphi $ sheet. We trace the states from an arbitrary starting point (stars) as they evolve from the virtual to bound states (circles) to resonances (squares). Insets $A$ and $B$ show the behavior of poles near the three-body threshold. Figure adapted from Ref.~\citep{Dawid:2023kxu}.}
    \label{fig:2}
\end{figure}

In the model under study, depending on the value of the two-body scattering length, $a$, the three-body amplitude has one or more poles on the physical and nearest unphysical Riemann sheets. Using techniques described in the previous section, we can freely access various sheets of the amplitude and explore the dependence of the pole positions on the interaction parameter, changing it from negative to positive infinity. We present the main result of our analysis in Fig.~\ref{fig:2}. Following Efimov~\citep{Efimov:1970zz, Efimov:1973awb}, we rephrase the binding energy of a given state, $\Delta E = E - 3m$, in terms of the (generalized) binding momentum, $\kappa = {\rm sign} (\re\Delta E)\sqrt{|m \re \Delta E| }$ and plot it against $1/ma$. We do it for the first three states (trimers) only. Some properties of the three-body spectrum are worth pointing out:
    \begin{enumerate}
        \item As observed most clearly on insets $A$ and $B$, energy levels of trimers are related by a simple rescaling. It is known as the Efimov phenomenon in the non-relativistic three-body problem and occurs for an infinite number of excited states (not shown). Although, due to the relativistic corrections, the ground state energy only approximately obeys this property, its behavior remains similar to that of the excited states. 
        \item The excited three-body resonances follow closed trajectories approaching the three-body threshold at two different values of $a$. They evolve from narrow near-threshold resonances to objects more closely resembling virtual states. Interestingly, we find that any two subsequent poles coincide at the threshold at the same value of this parameter. Our finding agrees with the non-relativistic investigation performed in Refs.~\citep{Bringas:2004zz, Hyodo:2013zxa}. \vspace{-2pt}
        \item The depicted evolution of excited resonances ends abruptly at these coincidence points. This behavior highlights the importance of the infinite number of Riemann sheets in a single-channel three-body problem. We claim that the trajectories of resonances extend to the higher sheets of the logarithmic three-body cut (not shown here), which in principle can be studied with contours of type $\Cc_3$ from Fig.~\ref{fig:1}. 
    \end{enumerate}
    \vspace{-2pt}

Most of these features are likely unique to the examined model. However, given that our parametrization of $\Mc_2$ is valid for any low-energy two-body scattering process and that the kinematic one-particle exchange between interacting pairs is a generic feature of the three-body scattering, it is conceivable that analysis of the more involved realistic systems features a spectrum qualitatively similar to the one explained here. Given the available knowledge of the non-relativistic three-body systems~\citep{Hammer:2010kp, Naidon:2016dpf}, we do not expect the emergence of the exact discrete scaling symmetry or perfectly cyclic trajectories in studies of realistic $\3 \to \3$ reactions. Nevertheless, the presented example of the three-body spectrum may serve as a tool guiding searches for three-body resonances and the state-of-the-art Lattice three-body calculations~\cite{Hansen:2020otl, Sadasivan:2021emk}.
\vspace{-5pt}

\section{Conclusions}
\label{sec:Conclusions}

In the trilogy of Refs.~\citep{Jackura:2020bsk, Dawid:2023jrj, Dawid:2023kxu} we have shown how to numerically solve and analytically continue the relativistic three-body integral equation derived from a generic EFT formulation and linked to a widely used finite-volume quantization condition~\citep{Blanton:2019vdk, Fischer:2020jzp, Hansen:2020otl, Blanton:2021llb, Draper:2023boj}. We paid special attention to describing our approach in a systematic and ready-to-implement manner. Similar methods will be required in future Lattice QCD analyses of the multi-body hadronic spectrum.

As explained in the text, the presented techniques rely on the integration contour deformation and careful analysis of the singularities of the constituents of the formalism. They allow one to extend the validity of the original equation to the complex energies and determine resonance poles of the $\3 \to \3$ amplitude. As a guiding example, we applied them to the simplest example of three-body interactions. We studied the evolution of the resulting spectrum across various Riemann sheets of the complex energy plane by varying the interaction strength. Within the infinite-volume fully relativistic framework, we recovered both the known non-relativistic~\citep{Efimov:1970zz, Hyodo:2013zxa} and finite-volume~\citep{Romero-Lopez:2019qrt} results.

Generalization of the analysis presented here to coupled-channel problems involving non-degenerate particles is underway. Given the conceptual development of the three-body infinite-volume techniques, and parallel advancement in their finite-volume counterparts, we believe the field of hadronic spectroscopy reached a stage where soon one will be able to perform a reliable lattice computation of the physical three-body states, such as $T_{cc}^+$ or $\chi_{c1}(3872)$.

\vspace{-5pt}

\section*{Acknowledgements}

The author acknowledges the financial support through the U.S. Department of Energy Contract no.~DE-SC0011637.

\bibliographystyle{JHEP}
\bibliography{main}

\end{document}